# Salt Contribution to the Flexibility of Single-stranded Nucleic Acid of Finite Length

Feng-Hua Wang[†], Yuan-Yan Wu[†], and Zhi-Jie Tan*

*Department of Physics and Key Laboratory of Artificial Micro- and Nano-structures of Ministry of Education, School of Physics and Technology, Wuhan University, Wuhan 430072, China*

Nucleic acids are negatively charged macromolecules and their structure properties are strongly coupled to metal ions in solutions. In this paper, the salt effects on the flexibility of single-stranded (ss) nucleic acid chain ranging from 12 to 120 nucleotides are investigated systematically by the coarse-grained Monte Carlo simulations where the salt ions are considered explicitly and the ss chain is modeled with the virtual-bond structural model. Our calculations show that, the increase of ion concentration causes the structural collapse of ss chain and multivalent ions are much more efficient in causing such collapse, and trivalent/small divalent ions can both induce more compact state than a random relaxation state. We found that monovalent, divalent and trivalent ions can all overcharge ss chain, and the dominating source for such overcharging changes from ion-exclusion-volume effect to ion Coulomb correlations. In addition, the predicted $Na^+$ and $Mg^{2+}$-dependent persistence length $l_p$'s of ss nucleic acid are in accordance with the available experimental data, and through systematic calculations, we obtained the empirical formulas for $l_p$ as a function of $[Na^+]$, $[Mg^{2+}]$ and chain length.

## I. INTRODUCTION

Nucleic acids have important biological functions in gene storage, transcription, and gene regulation. Their functions are strongly coupled to the structures and the proper structure changes.[1-12] Due to the polyanionic nature of nucleic acid backbone, the folding into compact native structures always involves strong Coulombic repulsions, thus requires metal ions in solutions, such as $Na^+$ and $Mg^{2+}$, to neutralize the negative backbone charges and stabilize the folded structures. Therefore, metal ions play essential roles in nucleic acid structures and functions.[1-12]

Single-stranded (ss) chain is an elementary structural and functional segment of nucleic acids. For example, RNAs structures generally consist of different type of ss loops, and ss chain is also the denatured state of nucleic acids.[13-21] Furthermore, ss chain is an important intermediate in many key biochemical processes, such as replication, recombination repair and transcription, and is specifically recognized by many proteins.[22] The flexibility of ss chain, which may be sensitive to ionic environment, plays a significant role in its interactions with other macromolecules, e.g., proteins.[23] Therefore, quantitative understanding how ionic condition, including ion concentration, ion valence and ion size, determines the flexibility of ss nucleic acids, is an important step toward understanding nucleic acid structures and functions. However, due to the negatively charged nature and strong dynamic conformational fluctuation, to quantitatively characterize the ion effects on the flexibility of ss nucleic acid chain is still a challenge, especially for long chains in multivalent ion solutions.

In general, there have been several classic polyelectrolyte theories for treating the ion-nucleic acid interactions: the counterion condensation (CC) theory,[24] the Poisson-Boltzmann (PB) theory,[25-27] and the Debye-Hückel (DH) theory.[28-31] These existed theories have been quite successful in predicting electrostatic properties of nucleic acids and proteins in ion/aqueous solutions.[24-31] However, the CC theory is based on the simplified line-charge structural model and is a double-limit law, thus is inapplicable to the conformational fluctuation of ss chain of finite-length.[24] The PB theory is a mean-field theory that ignores ion-ion correlations, which can be important for multivalent ions, e.g., $Mg^{2+}$.[32-34] The DH theory, characterized by a screened Coulomb potential between charged particles, is the linearized analytical form of PB equation, and thus applicable to the case of relatively weak electrostatic field and monovalent ion solution. Recently, to account for the effects of ion correlation and ion-binding fluctuation, a tightly bound ion (TBI) model was developed.[33] The model has been successful in predicting the $Na^+/Mg^{2+}$ effects in stabilizing DNA/RNA helices/hairpins,[18-20] and RNA tertiary structures.[35,36] However, for the ss nucleic acid with very strong conformational fluctuations, the TBI model at the present level is computationally very expensive.[18]

Parallel to the development of theories, computer simulations have been employed to the study of polyelectrolyte systems, which greatly enhanced our qualitative understanding of nucleic acid-ion interaction, since ss nucleic acid is a special one of polyelectrolytes and shares some general ion-dependent properties. However, until now, the simulational investigations were either based on different-level approximations or ignored specific nucleic acid structural features.[37-43] For example, most of them investigated macroscopic properties of bead-spring model of polyelectrolytes in salt-free or explicit dilute salt solutions, or predicted DNA structure changes by accounting for ion effects through a screened DH potential.[37-43] Beyond the coarse-grained simulation models, all-atom simulations can capture the molecule structures and solvent details at atomistic level.[44-47] But for ss chain with strong conformational fluctuation, all-atom simulations are computationally too expensive. Therefore, quantitative and systematic understanding on the flexibility of ss nucleic acids in ion solutions is still limited in both theories and computer simulations, especially for ss chains in multivalent ion solutions.[18]

In this work, we will employ the coarse-grained Monte Carlo simulations to systematically study the flexibility of ss nucleic acid chain of finite length in monovalent, divalent and trivalent salt solutions. Beyond the previous CC, PB and DH-based studies, the present method explicitly accounts for ion correlations and ion-binding fluctuations, and the structural model for ss nucleic acid is based on the near-realistic nucleic acid backbone derived from a variety of RNA structures in protein data bank (PDB). In the paper, we emphasize the ion-dependent macroscopic structural collapse and persistence length $l_p$, and the comparisons with the available experimental data. In addition, we obtain the empirical formulas for $l_p$ as a function of $[Na^+]$ and $[Mg^{2+}]$, and chain length $N$.

## II. MODEL AND METHOD
### A. Structural Model of ss Nucleic Acid

Our system consists of a ss nucleic acid and spherical ions dissociated from the added salt. The ions are considered explicitly and the ss nucleic acid is modeled as a coarse-grained structural chain with the virtual-bond model,[48-52] since the accurate all-atomistic representation would involve huge computational complexity and the simplified virtual-bond structural model can account for the backbone conformation while retain the advantages of a coarse-grained model.[48-52] As shown in Fig. 1a, a nucleotide is represented by the unit defined by two virtual bonds: $C_4$-P and P-$C_4$,[48-52] where P and $C_4$ stand for the phosphate and carbon ($C_4$) atoms, respectively. Each nucleotide carries a negative unit charge at the center of phosphate atom.[18] Phosphate and carbon atoms are treated as spheres with the respective van der Waals radii of 1.9Å and 1.7Å.[53] The distribution probabilities of bond lengths for $C_4$-P and P-$C_4$ and the bond angles for $C_4$-P-$C_4$ and P-$C_4$-P are both obtained through the statistical calculations over 40-individual of RNA structures (with the chain length $N$ ranging from 40 to 200-nt) deposited in PDB database, as shown in Figs. 1b and 1c; see Appendix for the PDB codes. Since the angles of $C_4$-P-$C_4$ and P-$C_4$-P are very similar, in the work, we do not distinguish the two angles. The previous studies have shown that the virtual-bond model could give good predictions on RNA secondary/pseudoknot structures and thermodynamics.[49-52]

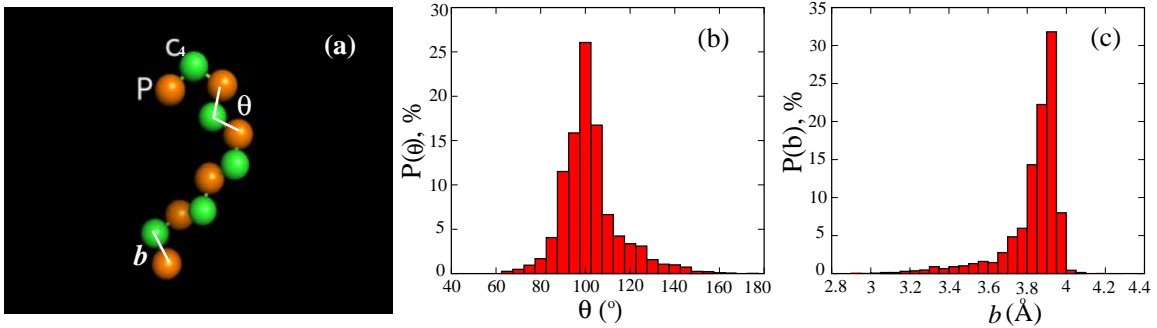

**FIGURE 1** (a) An illustration for the virtual-bond coarse-grained structure model for ss nucleic acid where a chain is represented by the sequential P (red color) and C4 (green color) atoms.[18,49-52] The two structural parameters, the bond angle $\theta$ and bond length $b$, define the backbone of the ss chain. (b, c) The normalized populations $P(\theta)$ and $P(b)$ of $\theta$ (b) and $b$ (c) for ss nucleic acid chain are obtained by the statistical analysis over the all-atom structures of 40 RNA molecules with the length ranging from 40-nt to 200-nt. The PDB codes of these RNAs are listed in the Appendix.

## B. Simulation Method

Monovalent, divalent, and trivalent salts are added into a cubic cell and dissociate into monovalent, divalent, and trivalent cations and coions respectively, which are modeled as charged spheres. Solvent (water) molecules are modeled implicitly as a medium with dielectric constant $\varepsilon$ ($\varepsilon$=78.3 at room temperature). In this study, the ions are assumed to be hydrated, and the radii of hydrated $Na^+$, $Mg^{2+}$ and $Co^{3+}$ ions are taken as 2.7Å, 3.6Å and 4.1Å,[54] respectively. The coions are treated as $SO_4^{2-}$ with hydrated radii 4.4Å[54] to decrease the simulation complexity (number of ions). The testing calculations show that the results for the coions of $Cl^-$ (~3.3Å[54]) are quantitatively similar to those for $SO_4^{2-}$ as coions. The simulational cell is always kept neutral and the periodic boundary condition is applied. To diminish the boundary effect, we always keep the cell size larger than a ss chain by 8 times of Debye length depending on the salt concentration. Our control tests by enlarging cell size (e.g., 12 times of Debye length) show that our results are rather stable.

Monte Carlo process is performed with the traditional Metropolis algorithm in a canonical ensemble.[55] To improve the sampling efficiency, two different types of moves are performed: conformations of ss chain are generated by pivot move and ions are moved by translation. The pivot move algorithm has been demonstrated to be rather efficient in sampling conformations of a polymer.[56,57] As shown in Fig. S1 (in Supporting Information), a ss chain can rapidly reach the equilibrium with the pivot move algorithm. In each run of the simulations, the first period of $0-3\times10^6$ Monte Carlo steps is used to relax the system to an equilibrium state, and the following $10^7$ steps are used to make statistical analysis on the structure properties of the system in equilibrium.

## C. Energy Functions

In the present model, four types of interactions are accounted for. The first one is the excluded-volume interaction applied for all the particles, including monomers (P and C4 atoms), cations, and coions, and can be given by a truncated Lennard-Jones potential

$$U_{LJ}(r) = \begin{cases} U_0\left[\left(\frac{\sigma}{r}\right)^{12}-\left(\frac{\sigma}{r}\right)^6\right], & \text{for } \frac{r}{\sigma} \leq 1; \\ 0, & \text{for } \frac{r}{\sigma} > 1, \end{cases} \quad (1)$$

where $r$ is the distance between the centers of two particles and $\sigma$ is the sum of their radii. The parameter $U_0$ is taken as 0.35 due to the soft H-atom exclusion from hydrated ions.[35,36,53]

The second energy is the electrostatic interactions between all charges (P atoms and ions)

$$U_{coul}(r) = k_B T l_B \frac{Z_i Z_j}{r}, \quad (2)$$

where $k_B$ is the Boltzmann constant, and $T$ is the temperature in Kelvin. $Z_i$ and $Z_j$ are the charges (in the unit of $e$) on the two particles $i$ and $j$. $l_B = e^2/(4\pi\varepsilon\varepsilon_0 k_B T)$ is the Bjerrum length. Here, $\varepsilon$ is the dielectric constant of solvent and $\varepsilon_0$ is the vacuum permittivity.

The third energy $U_{bond}$ is the bond connectivity potential energy used to describe the virtual bond length between the adjacent atoms (P and C4) along a ss nucleic acid backbone. The fourth energy $U_{angle}$ is the harmonic angle potential energy, which gives the virtual bond angle of ss chain. $U_{bond}$ and $U_{angle}$ are given by the following expressions respectively

$$\begin{aligned} U_{bond}(b) &= -k_B T \ln[P(b)]; \\ U_{angle}(\theta) &= -k_B T \ln[P(\theta)], \end{aligned} \quad (3)$$

where $P(b)$ and $P(\theta)$ are the normalized probabilities for bond length $b$ between neighboring (P and C4) atoms and for bond angle $\theta$ between the neighboring virtual bonds along the nucleic acid backbone, respectively. In the future work, the further improvement can be made on the $U_{bond}$ and $U_{angle}$ to more accurately catch the specific local dynamics and structures of RNAs in the all-atomistic model.[58] As shown in Fig. 1 and subsection of "Structural Model of ss Nucleic Acid", the two interaction potentials can give a good description for ss nucleic acid backbone.

## III. RESULTS AND DISCUSSIONS

In this section, we first investigate the ion-dependent structural collapse and ion-binding properties for ss nucleic acid chains of different lengths immersed in $Na^+$, $Mg^{2+}$, and $Co^{3+}$ salt solutions. Afterwards, we examine the ion size effect, and calculate the persistence length of ss nucleic acid and make comparisons with the available experimental data. Finally, we drive the empirical formulas for $l_p$ as a function of [$Na^+$] and [$Mg^{2+}$], and chain length, through the systematic calculations. Our calculations cover the broad ranges of ion concentrations: $[Na^+] \in [0.001M, 1M]$, $[Mg^{2+}] \in [0.03mM, 0.3M]$, and $[Co^{3+}] \in [0.01mM, 0.1M]$, and chain length range: $N \in [12\text{-nt}, 120\text{-nt}]$.

## A. Ion-dependent Collapse of ss Nucleic Acid Chain

Since the structural behaviors in monovalent, divalent and trivalent ion solutions share some general features, in the following, we first describe the general features for the ion-dependent structure collapse and then discuss the specific features for $Na^+$, $Mg^{2+}$ and $Co^{3+}$, respectively.

### 1. General features of structural collapse

As shown in Fig. 2 for $R_{ee}$ and Fig. S2 (in Supporting Information) for $R_{ee}/N$, ion-dependent collapse of ss chain shows the following general features for $Na^+$, $Mg^{2+}$, and $Co^{3+}$ salts:

(i) As ion concentration increases, $R_{ee}$ ($R_{ee}/N$) of ss chain decreases, which corresponds to the ion-induced structural collapse. Physically, the structural collapse of ss chain is opposed by the Coulomb repulsions between the backbone negative charges, while ions in solution can bind to nucleic acid and reduce such Coulomb repulsions. A higher ion concentration would reduce such repulsive force more strongly due to the lower entropy penalty for ion-binding and consequently stronger ion neutralization, causing the ion-induced collapse of the ss chain.

(ii) Longer ss chain shows the stronger ion-concentration dependent structural collapse. i.e., $R_{ee}$ ($R_{ee}/N$) decreases more sharply with the increase of ion concentration. Because longer chain involves stronger electric field in the vicinity, its structural collapse can cause stronger Coulomb repulsive force, consequently shows a stronger dependence on ion concentration.

In addition to the above described general behaviors, our calculations also show the special features on the structural collapse of ss chain in $Na^+$, $Mg^{2+}$ and $Co^{3+}$ solutions, respectively.

### 2. In $Na^+$ solutions

Fig. 2a and Fig. S2 (in Supporting Information) show that, the increase of $[Na^+]$ causes the structural collapse of ss chain from an extended state to a near random relaxation state at ~1M $[Na^+]$ where a ss chain behaves like a neutral chain, which is due to the stronger ion binding and the near-full-neutralization at very high $[Na^+]$.

### 3. In $Mg^{2+}$ solutions

$Mg^{2+}$ has higher charge as well as larger hydrated size than $Na^+$. As Fig. 2b shows, $Mg^{2+}$ is more efficient in inducing the structural collapse of ss chain than $Na^+$ beyond the mean-field concept such as ionic strength. For example, for $N$=48-nt ss chain, $R_{ee}$ at 0.03M $[Mg^{2+}]$ is close to that at 1M $[Na^+]$ (see Fig. 2). Moreover, the dependence of $R_{ee}$ on $[Mg^{2+}]$ is weaker than that on $[Na^+]$. Such specific role of $Mg^{2+}$ comes from the higher ionic charge of $Mg^{2+}$. Due to the stronger $Mg^{2+}$-phosphate attraction, $Mg^{2+}$-binding is more enthalpically favorable and (effectively) less entropically unfavorable than $Na^+$. Consequently, $Mg^{2+}$-binding and $R_{ee}$ are less dependent on ion concentration. Such higher efficiency of $Mg^{2+}$ over $Na^+$ in causing the chain collapse is more pronounced for longer chains.

### 4. In $Co^{3+}$ solutions

To extend our analysis to higher valence ions, we examine the $Co^{3+}$ solution. As shown in Fig. 2c and Fig. S2c (in Supporting Information), $Co^{3+}$ is much higher efficient in causing ss chain collapse than $Mg^{2+}$ and $Na^+$, and such higher efficiency more pronounced for longer chain. Moreover, ss chains show much weaker ion concentration-dependence of $R_{ee}$ for $Co^{3+}$ than those for $Mg^{2+}$ and $Na^+$ due to the much stronger ion binding affinity and much stronger ion Coulomb correlations.

In addition, as shown in Fig. 2c and Fig. S2c, the $Co^{3+}$-dependence of chain collapse shows a V-shaped $R_{ee}$ curve against $[Co^{3+}]$, i.e., with the addition of $Co^{3+}$, $R_{ee}$ ($R_{ee}/N$) decreases at low $[Co^{3+}]$ (≤0.003M), while increases when $[Co^{3+}]$ exceeds a certain value (~0.003M). This reexpanding transition of ss chain is similar to the redissolution observed for DNA/polyelectrolyte aggregates and may be attributed to the strong overcharging of ss chain by $Co^{3+}$;[59-65] see the following subsection of "Ion binding and Overcharging". Moreover, as shown in Fig. 2c and Fig. S2c for $N$=72-nt, it is notable that for $[Co^{3+}] \in [0.3mM, 0.01M]$, $R_{ee}$ ($R_{ee}/N$) is smaller than that of a neutral ss chain, which suggests a like-charged attractive force among ss chain. The mechanism for such intra-chain attraction will be illustrated in the section of "Ion Size Effect".

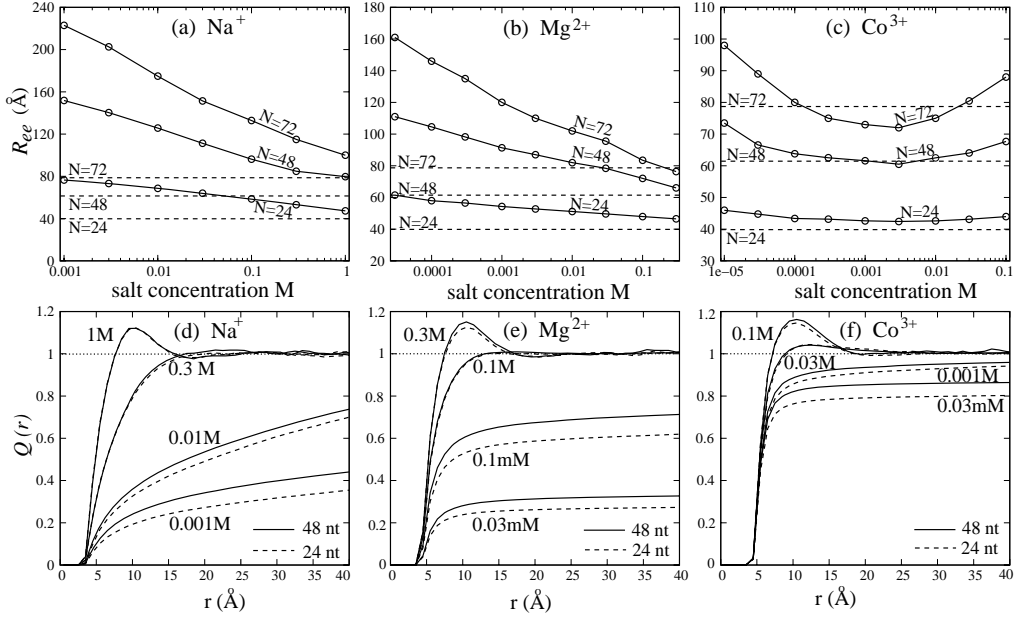

**FIGURE 2** (a-c) The end-to-end distance $R_{ee}$ of ss nucleic acid chain is shown as a function of $Na^+$, $Mg^{2+}$ and $Co^{3+}$ concentrations for different chain lengths $N$. Solid lines, the Monte Carlo simulations with explicit ions; Dashed lines, neutral ss chains as the references for the random relaxation state. (d-f) Integrated net charge distribution $Q(r)$ (Eq. 4) per nucleotide as a function of distance $r$ around ss nucleic acids ($N$=48-nt and 24-nt) at various salt concentrations. (d) $[Na^+]$=0.001M, 0.01M, 0.3M and 1M. (e) $[Mg^{2+}]$=0.03mM, 0.1mM, 0.1M and 0.3M. (f) $[Co^{3+}]$=0.03mM, 0.001M, 0.03 M and 0.1M.

## B. Ion Binding and Overcharging

### 1. General ion binding profiles

The above described macroscopic structural collapse is coupled to the microscopic ion-binding properties. We calculate the net charge distribution function $Q(r)$ which corresponds to the total net ion charges per nucleotide within a distance $r$ from the ss chain:

$$Q(r) = \int_{<r} \sum_\alpha Z_\alpha d^3\mathbf{r}, \quad (4)$$

where $Z_\alpha$ denotes the valence of $\alpha$ ion species. As shown in Figs. 2(d-f), $Q(r)$ monotonically increases and tends toward 1 as $r$ increases. When salt concentration is increased, the stairlike behavior of $Q(r)$ indicates more ions binding in the vicinity of chains. Such enhanced ion-binding is attributed to the lowered ion-binding penalty, and is responsible for the ion-induced structural collapse of ss chain described above. Also shown in Fig. 2, $Q(r)$ for $N$=48-nt is (slightly) higher than that for $N$=24-nt at the same ion conditions, i.e., each nucleotide of longer chain induces more binding ions, which is attributed to the stronger electric field in the vicinity of longer chain.

### 2. Overcharging

Interestingly, in the ion-binding profiles $Q(r)$ at high $[Na^+]$ (~1M), $[Mg^{2+}]$ (~0.3M) and $[Co^{3+}]$ (≥0.03M), $Q(r)$ can be apparently larger than 1, which suggests that ss chain is "overcharged" by $Na^+$, $Mg^{2+}$ and $Co^{3+}$, i.e., the binding ions are more than those can exactly neutralize the charges of a ss chain. Previous studies have suggested the two basic mechanisms responsible for overcharging: electrostatic correlation and the combination of excluded volume correlations and the electrostatic correlation.[66,67] In the present model, the salt ions are accounted for explicitly and the predicted "overcharging" should be related to the explicit ion properties (charge and size) which are ignored in the mean-field approximation.

To distinguish the driving force for such overcharging, we make the additional calculations by decreasing ion size for $Na^+$, $Mg^{2+}$ and $Co^{3+}$, respectively. As shown in Fig. 3a, the decrease of monovalent ion size can diminish/weaken the degree of overcharging, and when the ion radii are equal to 0.5Å, the overcharging disappears. Thus, the overcharging by $Na^+$ should mainly come from the ion excluded volumes. As suggested in previous studies for charged colloid system,[66,67] the more bulky cations increase (decrease) the total ion excluded volume (ion translational entropy), which decreases the entropic penalty for the bulky ions binding to ss chain.[66,67] Moreover, enlarging the ion size would reduce the formation of cation-coion pairs and high-order ion clusters at high salt concentrations[68,69]. For the case, even weak Coulomb correlations can drive the ion binding to ss chain and lead to the formation of a strongly correlated liquid along ss chain, where the overcharged state is energetically favorable.[66] However, excluded volume alone can never lead to notable overcharging at low ion volume fraction, since in the limiting case (uncharged nucleic acid and uncharged ions) coions and counterions have the same radial distribution.[66]

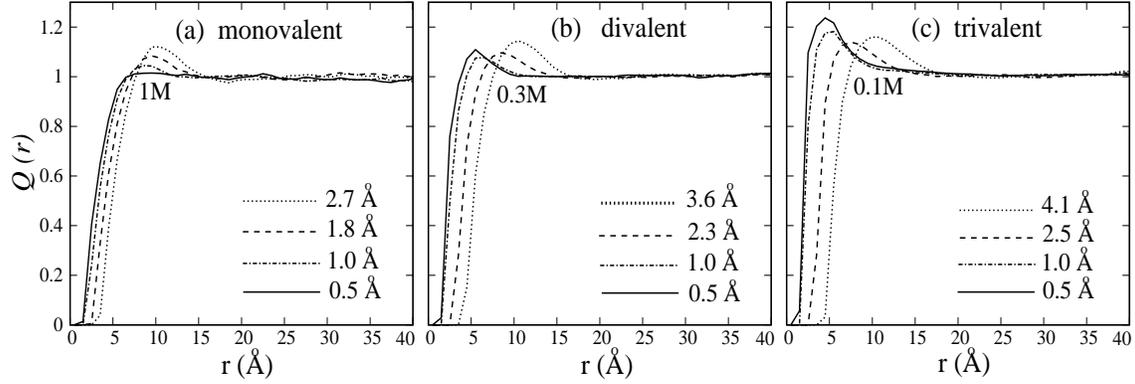

**FIGURE 3** Integrated net charge distribution $Q(r)$ (Eq. 4) per nucleotide as a function of distance $r$ from a 24-nt ss nucleic acid chain for four different ion sizes. (a) 1M monovalent salt with cation radii of 2.7Å, 1.8Å, 1.0Å, and 0.5Å. (b) 0.3M divalent salt with cation radii of 3.6Å, 2.3Å, 1.0Å, and 0.5Å. (c) 0.1M trivalent salt with cation radii of 4.1Å, 2.5Å, 1.0Å, and 0.5Å.

As shown in Fig. 3b for 0.3M $Mg^{2+}$, with the decrease of ion size, the overcharging is only slightly weakened. Even for the ion size of 0.5Å, $Q(r)$ is still apparently larger than 1. Therefore, the overcharging by divalent ions might be mainly due to the ion Coulomb correlations. As suggested in previous studies,[59-61] ion Coulomb correlations allow ions to self-organize to achieve low-energy state and consequently stronger ion binding, which could cause the overcharging. In addition, ion-exclusion-volume effect may also contribute to the overcharging by $Mg^{2+}$, which is responsible for the slight weakening in overcharging as ion size decreases.

For 0.1M $Co^{3+}$, Fig. 3c shows that the decrease of ion size can enhance the overcharging, suggesting that the overcharging by $Co^{3+}$ should mainly come from the ion-ion Coulomb correlations rather than ion-exclusion-volume effect, as compared to the $Na^+$ and $Mg^{2+}$ salts. Our predictions for the overcharging for the monovalent, divalent and trivalent ion solutions are also in agreement with the overcharging diagram for the planar electrical double layer (EDL) through an integral equation theory.[70] It shows that, for the case of weak ion-charged wall Coulomb interactions (similar to our $Na^+$ instance), a decreasing behavior of the number of adsorbed cations is predicted with the decrease of the ion volume fraction, which, in turn implies overcharging is dominated by ion excluded volume correlations, but the former decreasing trend of the number of adsorbed cations *versus* ion volume fraction is reversed for the case of strong ion-charged wall Coulomb interactions (similar to our $Co^{3+}$ instance) where Coulomb correlations dominate the overcharging.

Due to the high ionic charge of $Co^{3+}$, the strong overcharging by $Co^{3+}$ comes from the strong ion-ion Coulomb correlations, and the Coulomb repulsion between "overcharged" ss chain would contribute to the reexpansion of ss chain at high $[Co^{3+}]$ as concerned above and suggested in previous studies.[59,71] Nonetheless, as shown in Fig. 2b, the apparent reexpansion of ss chain is not observed where ss chain is overcharged by $Mg^{2+}$ at 0.3M $[Mg^{2+}]$. When the overcharging becomes stronger, i.e., $[Mg^{2+}]$ is increased to ~0.6M, the apparent reexpansion of ss chain can be observed (for $N$=48-nt, $R_{ee}$~70.1Å at 0.3M and ~73.7Å at 0.6M $[Mg^{2+}]$), which is consistent with Muthukumar *el al.*'s prediction for polyelectrolyte gel that the chain swells again when the overcharging occurs to a larger extent in divalent salts.[72] This is because, as discussed above, unlike $Co^{3+}$, the overcharging by $Mg^{2+}$ results from both of the ion Coulomb and ion-exclusion-volume correlations, while the reexpansion of ss chain would rely strongly on the sufficient overcharging by ion Coulomb correlations, e.g., >0.3M $[Mg^{2+}]$.

In this work, the radii of cations and coions are taken as the values estimated from experiments.[54] The overcharging of ss chain is related to coion size. The decrease of coion size would weaken the degree of the overcharging (we have examined the values of coion radii of 4.4Å, 3.5Å, 2.5Å and 1.0Å; data not shown). This is because smaller coions would interact more strongly with cations to likely form cation-coion pairs/clusters, thus could decrease the effective charge of cations and consequently weaken the overcharging.[68]

**C. Ion Size Effect**

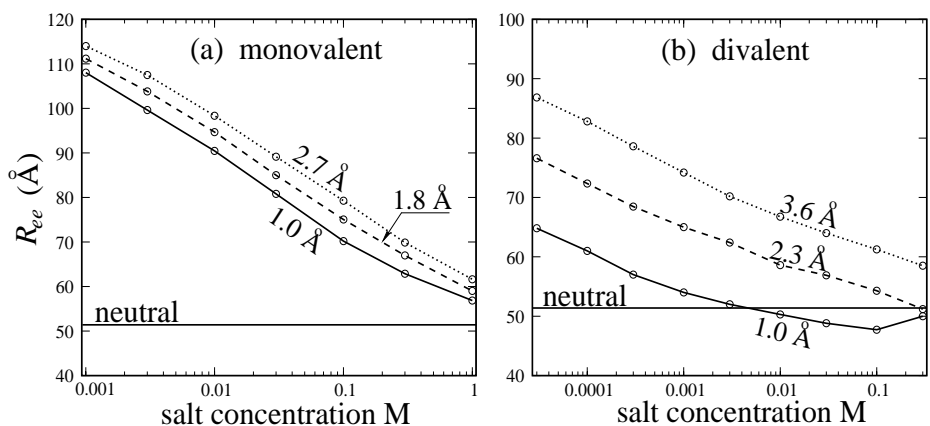

**FIGURE 4** The end-to-end distance $R_{ee}$ of a 36-nt ss nucleic acid chain for monovalent and divalent salts. (a) The monovalent cation radii of 2.7Å (dotted line), 1.8Å (dashed line) and 1.0Å (solid line). (b) The divalent ion cation radii of 3.6Å (dotted line), 2.3Å (dashed line) and 1.0Å (solid line).

In this subsection, we examine the ion size effect on the structural collapse of ss nucleic acid chain. In the calculations, the radii of monovalent cations are varied from 2.7Å to 1.0Å and the radii of divalent cations are varied from 3.6Å to 1.0Å. As shown in Fig. 4, the ion concentration-dependence of $R_{ee}$ shows qualitatively similar behaviors for ions with different radii.

For monovalent ion solution, the decrease of ion size causes the decrease of $R_{ee}$ and $R_{ee}$ becomes closer to that of the neutral chain at 1M. This is because small ions have stronger binding affinity and are more efficient in charge neutralization. Such enhanced ion-binding by decreased ion size is shown in Fig. S3a (in Supporting Information). At high (1M) ion concentration, ss chain is overcharged, which is mainly attributed to the bulky ion size and such overcharging would disappear when monovalent ion sizes become very small, as discussed above.

For divalent ions with different sizes, as shown in Fig. 4b, $R_{ee}$ shows the similar ion-concentration dependence. For the ion radii 3.6Å and 2.3Å, the chain is still less compact than a neutral chain. But for ions with radii of 1.0Å, $R_{ee}$ can decrease below the value of neutral chain when ion concentration exceeds ~0.01M, which suggests an intra-chain attractive force. The interesting phenomenon of the intra-chain attractive force is related to the ion-bridge configuration which has been suggested in experiments[40] and theory[62], where cations orderly adsorb between approaching negatively charged groups like London attraction force in hydrogen molecule.[73] To directly illustrate the microscopic structure for the ion-bridge along chain, we present two typical snapshots. As shown in Fig. 5a, at low divalent cation concentration, only a small amount of ions condense on a ss chain and the chain exhibits an extended state. At high ion concentration (shown in Fig. 5b), on the other hand, much more cations bind to the chain and the chain becomes a relatively compact state. It is clearly shown that, along the chain, many configurations of phosphate-ion-phosphate are formed, and such ion-bridging-induced attractive force can overcome the chain intrinsic stiffness to cause the state with smaller compactness ($R_{ee}$) than the random relaxation state.

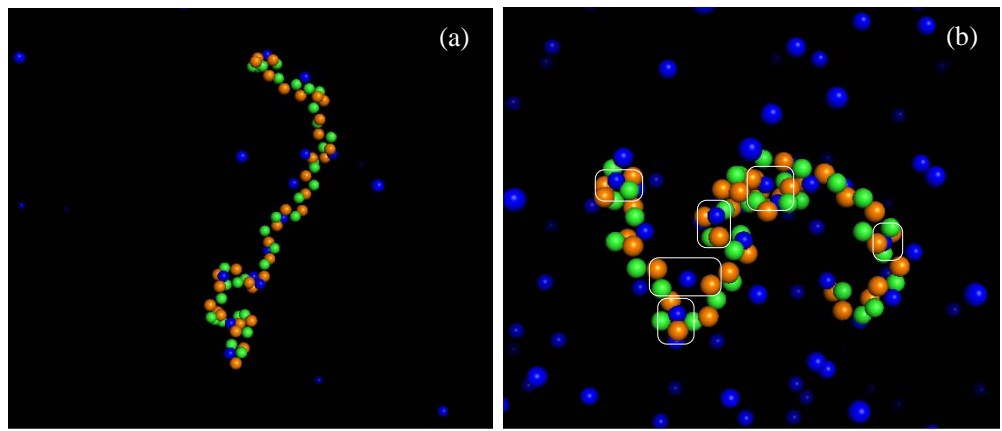

**FIGURE 5** Snapshots for a 36-nt ss nucleic acid in divalent salt solutions with cation radii of 1Å. (a) The chain in an extend conformation at 0.1mM divalent salt. (b) The chain in a condensed conformation at 0.1M divalent salt. The ss chain is represented by a sequence of P and $C_4$ atoms which are displayed in red and green colors, respectively. The divalent cations are shown in blue color, and coions are not shown in the snapshots. The ion-bridging configurations of P atom-divalent ion-P atom are marked with white rectangular boxes.

### D. Persistence Length $l_p$ versus $[Na^+]$ and $[Mg^{2+}]$

*1. Predicting $Na^+$ and $Mg^{2+}$-dependent persistence length*

Based on the conformational ensemble of ss chain in equilibrium, we calculate the ion-dependent persistence length $l_p$ of ss nucleic acids of different lengths, since $l_p$ quantitatively characterizes the flexibility of a ss chain.[74-78] The persistence length can be estimated in different ways based on the simulation data[74-79]. Here, the persistence length $l_p$ was calculated as the projection length[71,75,77]

$$l_p = \frac{1}{b} \sum_{i=1}^{N_m-1} <\vec{r}_1 \cdot \vec{r}_i>, \quad (5)$$

where $b$ is the bond length and $<...>$ denotes the ensemble average. $\vec{r}_i$ is the bond vector and $N_m$ is the number of monomers (there are $N_m-1$ bonds). Here, $l_p$ is approximated as the average projection of the end-to-end vector for the subchain from bond $i$ to the last bond onto the direction of the first bond.[71,75,77] In addition, we also use another method to estimate $l_p$ based on the worm-like model,[78,80]

$$R_g^2 = (\frac{Ll_p}{3}) - l_p^2 + (\frac{2l_p^3}{L}) - (\frac{2l_p^4}{L^2})(1-e^{-L/l_p}), \quad (6)$$

where $L=(N_m-1)*b$ is the contour length, and $R_g$ is the radius of gyration of ss chain. As shown in Fig. S4 (in Supporting Information), the two methods (Eq. 5 and Eq. 6) give the similar predictions.

Fig. 6 shows the predicted $l_p$ as a function of $[Na^+]$ and $[Mg^{2+}]$ for ss nucleic acids of different chain lengths. When ion concentration is increased, $l_p$ decreases from a high value and converges to low value of ~8Å which is nearly independent of chain length $N$. Fig. 6 also shows that longer ss chain has larger $l_p$ at low ion concentration and the stronger dependence on ion concentration. Higher ion concentration leads to stronger ion-binding, causing the decrease of $l_p$ and higher flexibility. At very high ion concentration, the chain would get nearly full-neutralized, thus $l_p$ converges to a low value which results from the intrinsic flexibility of ss nucleic acid.[81-84] From the trend of $l_p$ versus chain length $N$, we can expect that $l_p$'s will approach to the values of infinite-length chain as $N$ exceeds 120-nt.

*2. Comparisons with experimental data*

As shown in Fig. 6, the predicted $l_p$'s are in reasonably good agreement with the available experimental data for both of $Na^+$ and $Mg^{2+}$ solutions and fall within the rather large range from 6.4 to 65Å obtained by a number of different experiments. Tinland *et al.* experimentally found $l_p$'s of ss calf thymus DNA to be around 8(±2.5)Å, 25.8(±4.2)Å, and 49.4(±15.5)Å at 0.1M, 0.01M and 0.001M $[Na^+]$ respectively,[81] which are in accordance with the predicted $l_p$'s (~11.5Å, 23.5Å and 45Å for 120-nt, respectively) except that the measured value is slightly higher than our prediction at 0.001M $[Na^+]$. Such slight discrepancy is because the chain length in our calculations is finite ($N≤120$-nt) and the chain length effect is strong only at low salt concentration. Kuznetsov *et al.* used equilibrium DNA hairpin melting profiles to obtain $l_p$ of 14Å for poly(dT) strands at 0.1M $[Na^+]$,[82] which is in good agreement with our value (~12Å for 120-nt). Mechanical stretching of ss λ-DNA gave $l_p$ around 8Å in a 0.15M $[Na^+]$ solution,[83] which is also close to our prediction (~10Å for 120-nt). Recently, the $Na^+$-dependent $l_p$'s obtained through small-angle x-ray scattering by Pollack *et al.* for $dT_{40}$ are also close to our predictions except for slight deviation of ~3Å.[84] $l_p$'s for a 20-bp RNA hairpin measured by force spectroscopy method are in good agreement with our predictions.[85]

In contrast to $Na^+$ solutions, only few experiments were performed to get $l_p$ in pure $Mg^{2+}$ solutions. A mechanic study found $l_p$~6.7Å for denatured ss DNA at 0.05M $[Mg^{2+}]$.[86] Rivettie *et al.* estimated $l_p$~13Å for ~400-nt ss DNA in 0.002M $[Mg^{2+}]$ solution with 0.01M $[Na^+]$.[87] These experimental values are close to our values ($l_p$~8.6Å at 0.05M $[Mg^{2+}]$ and ~11.4Å at 0.002M $[Mg^{2+}]$ for 120-nt). Recently, Pollack *et al.*[78] obtained $Mg^{2+}$-dependent $l_p$'s for $dT_{40}$ in a 0.02M Tris buffer. It is known that 0.01M of monovalent ions can dissociate from 0.02M Tris buffer and can counteract the efficient role of $Mg^{2+}$.[88] With the decrease of $[Mg^{2+}]$, the experimental $l_p$ increases from 8Å at 0.05M $[Mg^{2+}]$ ($Mg^{2+}$-dominating case) to 19Å at 0.1mM $[Mg^{2+}]$ (monovalent ion-dominating case). The $l_p$'s at the $Mg^{2+}$- and monovalent ion-dominating cases agree with our values ($l_p$~8.3Å at 0.05M $[Mg^{2+}]$ and ~17.7Å at 0.01M $[Na^+]$ for 40-nt). Similarly, Ritort *et al.* recently measured $Mg^{2+}$-dependent $l_p$'s of a 20-bp RNA hairpin in a 0.1M Tris buffer.[85] As $[Mg^{2+}]$ decreases from 0.01M to 0.01mM, $l_p$ increases from ~7.5Å to ~15Å. The $l_p$'s at the two limiting cases also agree with our values ($l_p$~9.2Å at 0.01M $[Mg^{2+}]$ and $l_p$ ~12.3Å at 0.05M $[Na^+]$ for 40-nt). Certainly, the overall quantitative comparisons with the experimental data of Pollack et al.[84] and Ritort et al.[85] require the further theoretical modeling for ss chain in mixed $Na^+$/ $Mg^{2+}$ solutions.

The deviations from different experiments reflect, to some extent, the variation of persistence length values on ionic strength, chain sequence, secondary structure and experimental techniques. Overall, our results reasonably agree with previous measurements that utilized different techniques.

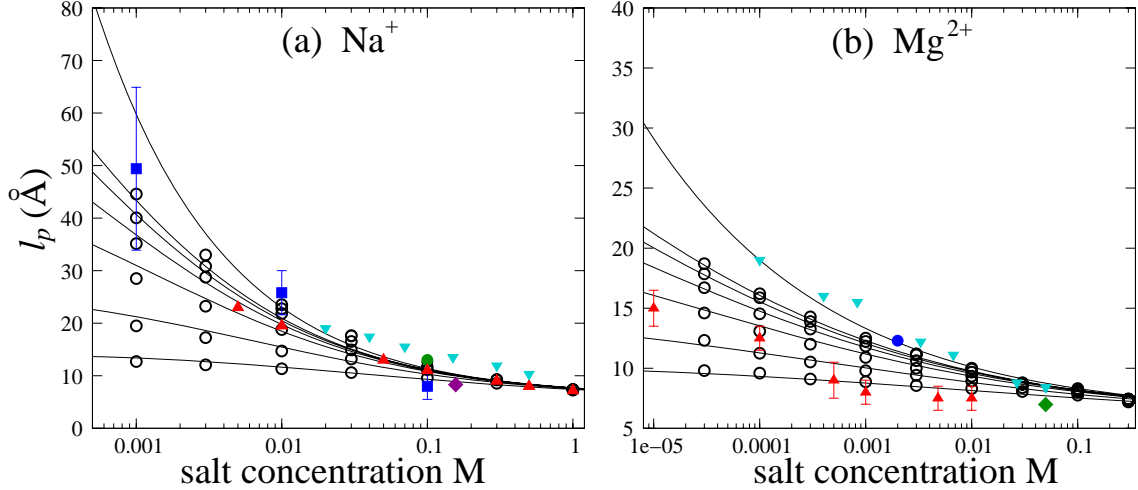

**FIGURE 6** The persistence length $l_p$'s of ss nucleic acid chain as functions of (a) $Na^+$ and (b) $Mg^{2+}$ concentrations for different chain lengths. Open circles, data from the Monte Carlo simulations; Solid lines, the empirical relation Eq. 8 for the ss chain of different lengths (from top to bottom, $N=\infty$, 120, 96, 72, 48, 24, 12-nt); Filled symbols, experimental data. (a) Blue ■, a ss fragment of calf thymus DNA in $Na^+$ solutions;[81] Green ●, poly(dT) strands in a $Na^+$ solution;[82] Magenta ♦, ss λ-DNA in a $Na^+$ solution;[83] Cyan ▼, $dT_{40}$ in $Na^+$ solutions with 0.02M Tris buffer;[84] Red ▲, a 20-bp RNA hairpin in $Na^+$ solutions with 0.1M Tris buffer.[85] (b) Cyan ▼, $dT_{40}$ in $Mg^{2+}$ solutions with 0.02M Tris buffer;[84] Red ▲, a 20-bp RNA hairpin in $Mg^{2+}$ solutions with 0.1M Tris buffer;[85] Green ♦, a ss DNA in a $Mg^{2+}$ solution;[86] Blue ●, a ~400-nt ss DNA in a $Mg^{2+}$ solution with 0.01M $Na^+$.[87]

### 3. $Na^+$ versus $Mg^{2+}$ on $l_p$

Besides the above described similar trend, $l_p$ shows apparently weaker dependence on $[Mg^{2+}]$ than on $[Na^+]$ and $Mg^{2+}$ is more efficient in achieving the same $l_p$ than $Na^+$. For example, for a 48-nt chain, 0.01M $Mg^{2+}$ and 0.3M $Na^+$ can achieve the same $l_p$ (~9.3Å). In addition, $l_p$ at high $[Mg^{2+}]$ is slightly lower than that at high $[Na^+]$. For example, $l_p$~7Å at 0.3M $Mg^{2+}$ is less than that $l_p$~8Å at 1M $Na^+$, which is in accordance with the experiments ($l_p$~8.4Å in 0.05M $[Mg^{2+}]$ and $l_p$~10.5Å in 0.5M $[Na^+]$)[78]. Such higher efficient role of $Mg^{2+}$ comes from the higher valence of $Mg^{2+}$ than $Na^+$ as discussed above.

### 4. Empirical formulas for $l_p$

From previous studies,[81-87] the persistence length of nucleic acid can come from two contributions: an intrinsic contribution $l_p^0$ which results from the intrinsic rigidity of ss chain and may be sequence-dependent, and an electrostatic contribution $l_p^e$ which is due to electrostatic interaction and is dependent strongly on the ion environment and chain length[81-87]:

$$l_p = l_p^0 + l_p^e . \qquad (7)$$

Our calculations allow us to systematically calculate $l_p$'s of ss nucleic acids of different lengths for various $[Na^+]$ and $[Mg^{2+}]$. Based on the calculations, we can obtain the following empirical expressions for $l_p$

$$l_p = l_p^0 + \frac{1.7}{[Na^+]^{0.5} + 1.6/(N-4)}, \quad \text{for } Na^+;$$

$$l_p = l_p^0 + \frac{1.3}{[Mg^{2+}]^{0.25} + 3.5/N}, \quad \text{for } Mg^{2+}, \qquad (8)$$

where $[Na^+]$ and $[Mg^{2+}]$ are in molar, and $l_p^0$ ~6Å for generic (random) sequences. Our result of the $l_p$~$[Na^+]^{0.5}$ for long chain is consistent with the Barrat-Joanny theory ($l_p$~$[Na^+]^{-0.5}$) for flexible polyelectrolytes,[89] while the dependence on $[Na^+]$ is weaker than that from the Odijk-Skolnick-Fixman theory ($l_p$~$[Na^+]^{-1}$) for semiflexible polyelectrolytes[28,90]. Furthermore, the salt-dependence becomes weaker for shorter sequence. According to the Odijk-Skolnick-Fixman theory and Barrat-Joanny theory, where $l_p$ consists of an intrinsic contribution and a contribution caused by electrostatic repulsion, $l_p$ of the double-stranded DNA is nearly independent of ion concentration at high monovalent salt.[28,90] In contrast, a theory proposed by Manning indicates that electrostatic and nonelectrostatic persistence lengths are not additive, and the dominant role in double-stranded DNA stiffness comes from the electrostatic repulsion of DNA charges.[91] At the meanwhile, through the simulations, Savelyev *et al.* predicted that a fine balance of electrostatic and elastic effects on $l_p$ and favored Manning's approach over Odijk-Skolnick-Fixman theory at high monovalent salt for double-stranded DNA.[92,93] While for ss nucleic acid chain studied here, $l_p$ decreases apparently with the increase of [Na$^+$] until 0.1M, and above 0.1M, $l_p$ continues to decrease slightly until 1M. Moreover, $l_p$ of ss chain has a weaker ion concentration-dependence than that of double-stranded DNA. For convenience, following the previous works on ss nucleic acid chain,[60,71,78,81,84] we use the above functional forms (Eq. 8) to derive the empirical formulas for $l_p$. As shown in Fig. 6, the above expressions for $l_p$ give good fits to our predictions and the available experimental data.[81-87] Such parameterized empirical formulas may be practically useful.[18,81]

## IV. CONCLUSIONS

In the present work, we have employed the Motel Carlo method to systematically and explicitly investigate ion effects on the flexibility of single-stranded nucleic acids of different lengths. The study covers the effects of ion concentration, ion valence and ion size. The major findings are in the following:

(i) The addition of Na$^+$ would induce ss chain to collapse from an extend state at low [Na$^+$] to a near-random relaxation state at high [Na$^+$] (~1M).

(ii) Multivalent ions are more effective than Na$^+$ in inducing the structural collapse of ss chain, and small divalent/trivalent ions can cause more compact state than the random relaxation state.

(iii) At high ion concentration, ss nucleic acids can be overcharged by Na$^+$, Mg$^{2+}$, and Co$^{3+}$. The overcharging in Na$^+$ and in Co$^{3+}$ solutions are dominated by the ion-exclusion-volume effect and ion-ion Coulomb correlations, respectively. For Mg$^{2+}$, both of ion-ion Coulomb correlation and ion excluded volume contribute to the overcharging.

(iv) Our predicted persistence lengths of ss nucleic acids agree well with the available experimental data, and we derive the empirical formulas for the persistence length as a function of [Na$^+$] and [Mg$^{2+}$], and the chain length.

Although the present work gives reasonable predictions and the predicted persistence length agree with the available experiment data, the present model also involves the following important approximations.

First, in the work, the solvent (water) molecules are implicitly approximated as the continuous medium with high dielectric constant, since explicit solvent model would involve huge computation complexity and limit the sampling on chain conformational ensemble. Such approximation has been validated in various previous calculations on salt effects in nucleic acid thermodynamics.[38,41]

Second, the ss nucleic acid is modeled as a coarse-grained chain with a virtual-bond model, since the all-atom structural model is computationally too complex. Such simplification is not a nonphysical approximation due to the distribution of nucleic acid charges on its backbone and the long-ranged nature of electrostatics, and the simplification has been validated in describing hairpin formation in ion solution.[18] Moreover, the two potentials of bond length and bond angle in our simulation are obtained by Boltzmann inversion of the corresponding distribution functions collected from PDB structures. Further adjustments accounting for cross-correlations should be included for retaining sufficient fidelity to the RNA fully atomistic dynamics.[58]

Third, the ions are assumed to be hydrated and the effects of ion dehydration and specific binding are ignored. For the ss chain with moderate charge density on backbone, the ion dehydration effect might be weak. Although the inner sphere complexation between cations such as Mg$^{2+}$ and polynucleotide is weak,[94] the ionic dehydration effect may become important for the ion-binding near nucleic acid surface at the high salt concentration. As shown above, the overcharging of ss chain has an ion size-dependence (see Fig. 3). Thus, to take into account short-range hydration effect is necessary in the future work on overcharging, referring to all-atom molecule dynamic simulations and other more accurate coarse-grained models. It is worth mentioning that Savelyev and Papoian have proposed a coarse-grained modeling method for short-range hydration effect. The incorporation of such effect in modeling double-stranded DNA in ion solutions has shown the accurate reproduction of atomistic behavior of ions.[58,95]

Finally, the present structural model of ss chain ignores the bases and thus cannot treat the sequence effects such as base-pairing/stacking and self-stacking. Previous studies have shown that, some specific sequences (e.g., poly (A) and poly (C)) would exhibit strong self-stacking and can form ss helices, while other most sequences (e.g., poly (U) and poly (T), and generic ss DNA) behave like coils.[1] Nevertheless, the present work forms an important step towards a more complete coarse-grained model to account for important sequence effect which directs the nucleic acid folding,[96-99] and is helpful for understanding the salt-dependent macroscopic flexibility of single-stranded nucleic acids of finite-length.

## APPENDIX

### RNA Structures

As described in "Model and Method", we calculate bond length and bond angle distributions with the use of the RNA structure data stored in Protein Data Bank (http://www.rcsb.org/pdb/home/home.do). The PDB codes of RNA chain structures used in our analysis include 3E5F, 2K4C, 3LA5, 1P5O, 1P5P, 1S9S, 1U8D, 1YM0, 1Z43, 2B57, 2G9C, 2EES, 3DIL, 3DIM, 3DIQ, 3DIR, 3DIS, 3DIX, 3DIZ, 3DIY, 3DJ0, 3DJ2, 3G4M, 3GX3, 3FO4, 3GX2, 2L0U, 1XJR, 1KXK, 2GIS, 3D0U, 3D0X, 3A3A, 3G0G, 1EVV, and 3F04. These individual RNA molecules contain the number of nucleotides ($N$) in the range 40-nt<$N$<200-nt.

## ASSOCIATED CONTENT

### Supporting Information

Gyration of radius $R_g$, end-to-end distance $R_{ee}$, net ion charge distribution $Q(r)$ and persisten length $l_p$ for ss nucleic acid chain.

## AUTHOR INFORMATION

### Corresponding Author

*E-mail: zjtan@whu.edu.cn

### Author Contributions

[†]These authors contributed equally.

### Notes

The authors declare no competing financial interest.

## ACKNOWLEDGMENTS

We are grateful to Shi-Jie Chen, Wenbing Zhang, Song Cao, and Huimin Chen for valuable discussions. This work was supported by the National Science Foundation of China grants (10844007, 11074191, and 11175132), the Program for New Century Excellent Talents (NCET 08-0408), the Fundamental Research Funds for the Central Universities (1103007), the National Key Scientific Program (973)-Nanoscience and Nanotechnology (No. 2011CB933600) and by SPF for ROCS, SEM. One of us (Y.Y.W) also thanks financial supports from the interdisciplinary and postgraduate programs under the "Fundamental Research Funds for the Central Universities".